\documentstyle[12pt]{article}
\begin{document}
\large

\newpage
\begin{center}
{\bf THE UNITED THEORY OF THE TWO FIELDS OF THE ELECTRIC
AND MAGNETIC NATURE}
\end{center}
\vspace{1cm}
\begin{center}
{\bf Rasulkhozha S. Sharafiddinov}
\end{center}
\vspace{1cm}
\begin{center}
{\bf Institute of Nuclear Physics, Uzbekistan Academy of Sciences,
Tashkent, 702132 Ulugbek, Uzbekistan}
\end{center}
\vspace{1cm}

Each of electrically charged particles testifies in favor of the existence
of a kind of the magnetically charged monoparticle. As a consequence,
only the corresponding mononeutrinos answer for quantization of the
electric charges of all the neutrinos. Therefore, to understand the
nature of matter at the fundamental level one must use the electromagnetic
field as the field of the unified system of the photon and monophoton where
the electric and magnetic forces of the nature are united. Some logical and
laboratory confirmations of the availability of compound structure of gauge
invariance have been listed which say also about the connection between
the states of elementary particles and monoparticles.

\newpage
A study of the behavior of electrons and their neutrinos in a nucleus
Coulomb field shows clearly that between the mass of a particle and its
charge there exists a sharp dependence \cite{1}. For the light Dirac
neutrino $(\nu=\nu_{e})$ it has the form \cite{2}
\begin{equation}
e_{\nu}^{E}=
-\frac{3eG_{F}(m_{\nu}^{E})^{2}}{4\pi^{2}\sqrt{2}}, \, \, \, \, e=|e|.
\label{1}
\end{equation}
Here $m_{\nu}^{E}$ and $e_{\nu}^{E}$ are the neutrino electric rest
mass and charge.

Such an intimate connection saying about the gravitational structure
of a Coulomb interaction and reflects the fact that the mass and charge
of a particle correspond to the most diverse form of the same regularity
of the nature of this field \cite{2,3}.

On the other hand, it is known that in the framework of the hypothesis of
field mass based on the classical theory of an extensive electron \cite{4},
a particle all the mass is purely electric. However, according to such
a point of view, the charge distribution of the neutrino will not be steady.

Our conclusion is that each of existing types of charges come forwards
in the system as the source of a kind of the inertial mass \cite{5}. Herewith
all the mass of the neutrino coincides with its united rest mass $m_{\nu}^{U}$
which includes in self the electric $m_{\nu}^{E},$ weak $m_{\nu}^{W},$ strong
$m_{\nu}^{S}$ and some other components:
\begin{equation}
m_{\nu}^{U}=m_{\nu}^{E}+m_{\nu}^{W}+m_{\nu}^{S}+....
\label{2}
\end{equation}

Such a compound structure giving the possibility to directly look at the
nature of matter is observed as a consequence of the availability of the
united charge $e_{\nu}^{U}$ in neutrino equal to its all the charge \cite{5}:
\begin{equation}
e_{\nu}^{U}=e_{\nu}^{E}+e_{\nu}^{W}+e_{\nu}^{S}+...,
\label{3}
\end{equation}
where the indices $E,$ $W$ and $S$ correspond to the electric,
weak and strong parts of charge.

Thus, it follows that if the neutrino possesses a Dirac mass then all
possible components of this mass establish the intraneutrino interratio
between the forces of the electric and unelectric nature. Therefore, the
charge distribution of the neutrino must be steady \cite{6}.

The purpose of the present article is to elucidate what the above - mentioned
duality of matter say about structures of electromagnetic field, gauge
invariance and charge quantization law.

In the framework of the loop phenomena, the neutrino must be electrically
neutral at the conservation of gauge invariance \cite{7}. It appears that
here on the basis of (\ref{1}) one can will decide a question about the
equality of the neutrino electric mass to zero.

But if $m_{\nu}^{E}=0,$ the neutrino is strictly Majorana particle \cite{5}.
At the same time the existence of massive Dirac neutrinos is by no means
excluded experimentally. Insofar as their neutrality in the loop approximation
is concerned, it reflects just a latent regularity of general picture of gauge
invariance which requires the study of the structure of electromagnetic field.

At first sight it appears that a connection between $e_{\nu}^{E}$ and
$m_{\nu}^{E}$ is incompatible with charge quantization. Such a regularity,
however, takes place owing to the magnetic monopoles. It is also known that
the standard $SU(2)_{L}$$\otimes$$U(1)$ theory, by itself, does not
require \cite{8} that the magnetic monopoles be exist in the form as was
suggested by Dirac \cite{9}. Therefore, from point of view of a massive
four - component neutrino, it should be expected that each of electrically
charged particles testifies in favor of the existence of a kind of
a <monoparticle> namely a particle having the magnetic mass and charge.
Then it is possible, for example, only the corresponding mononeutrinos
are responsible for quantization of the electric charges of all the neutrinos.

There exists a range of other phenomena, in which a fundamental part
is said of monoparticles. One of such systems may serve as the source
of electromagnetic field. But unlike earlier known, it must not be usual
gauge boson. The point is that the same photon does not lead to the
appearance of electromagnetic field. It possesses the electric mass \cite{10}
and charge \cite{11}.

From point of view of a Dirac neutrino, this gauge state will indicate to the
existence of the monophoton with the magnetic mass and charge. In other words,
the photon $\gamma_{E}$ and monophoton $\gamma_{H}$ must be source of the
electric $\vec{E}$ and magnetic $\vec{H}$ fields respectively. Under shuch
circumstances the electromagnetic field $(\vec{E}, \vec{H})$ appears as
the field of the unified system of the photon and monophoton $(\gamma_{E},
\gamma_{H})$ where the electric $\vec{F}_{E}$ and magnetic $\vec{F}_{H}$
forces of the nature are united.

To elucidate the compound structure of its gauge invariance one must apply
once more to the question whether the same particle has simultaneously both
electric and magnetic charges. There exist, however, many uncertainties both
in the nature and in the behavior of these types of charges. Nevertheless,
if we assume \cite{12} that the possibility of the existence of such Dirac
fermions is not excluded, the neutrino interaction with an electromagnetic
field of emission arises at the expense of exchange simultaneously both by
the photon and by the monophoton. The latter would lead us to the implication
that the invariance of these types of gauge bosons concerning C, P and T, and
also their combinations CP and CPT are not different.

Finally, insofar as the appearance of magnetic field in usual laboratory
conditions is concerned, to this one must apply as to one of the available
experimental data confirming the existence of monoparticles and saying that
any electrically charged particle can be converted into the corresponding
monoparticle and vice versa. Of course, these transitions and many other
aspects of compound structure of each of existing types of charges and
masses open of principle possibilities for creation at the fundamental
level of truly physical theories of naturally united gauge fields.

\end{document}